\documentclass{article}
\usepackage[preprint]{neurips_2025}
\usepackage{hyperref}
\usepackage{url}
\usepackage{booktabs}
\usepackage{amsmath, amssymb}
\usepackage{graphicx}
\usepackage{microtype}
\usepackage{xcolor}

\title{Neural Variable Name Repair: Learning to Rename Identifiers for Readability}

\author{
  Akshat Bagade \quad Maanas Baraya \quad Chittebbayi Penugonda \quad Muhammad Yousuf \\
  Georgia Institute of Technology \\
  \texttt{\{abagade6,mbaraya3,cpenugonda3,myousuf6\}@gatech.edu}
}

\begin{document}
\maketitle

\begin{abstract}
Developers routinely work with source files whose variable names are generic or misleading, and with teams moving quickly, many functions are left undocumented. This slows comprehension, increases the risk of subtle bugs, and makes it harder for both humans and large language models (LLMs) to reason about code. We study \emph{variable name repair}: given a real C++ function where all occurrences of one local or parameter name have been replaced by a placeholder (e.g., \texttt{<ID\_1>}), the goal is to generate a natural, descriptive replacement name.

We automatically construct this task from the C++ portion of BigCode’s \emph{The Stack} by parsing functions with Tree-sitter, masking a single identifier, and treating the original name as supervision. On top of Llama~3.1--8B, we build a pipeline with (i) warmup and dropout schedules for more stable fine-tuning, (ii) LoRA adapters for efficient specialization on identifier repair, and (iii) a dual-encoder reranker over top-$k$ generator candidates. We evaluate using exact match, Top-$5$ Hit, and an embedding-based partial similarity score (0--100) that gives credit for near-synonyms and format variants (e.g., \texttt{jsonValue} vs.\ \texttt{json}).

On a held-out set of 200 C++ functions, a zero-shot Llama~3.1 baseline reaches 6.1\% exact match. Our best LoRA-tuned model (with warmup and dropout) achieves 43.1\% exact match, 50.2\% Top-$5$ Hit, and an 82.03 partial-match score. A dual-encoder reranker further improves selection quality without modifying the underlying generator, suggesting that task-specific fine-tuning plus reranking is a promising approach for practical identifier repair tools.
\end{abstract}

\section{Introduction}

Readable code relies heavily on well chosen identifier names. When names are missing (\texttt{var1}), or misleading, developers spend more time reconstructing intent from control flow and types, and tools such as static analyzers, code search, and LLM based assistants work with degraded signal. In large codebases and legacy systems, renaming poor identifiers is often deferred or done inconsistently, even though it would benefit both human maintainers and automated tools.

We focus on a concrete version of this broader problem: \emph{variable name repair}. Given a single C++ function where one local or parameter identifier has been systematically replaced by a placeholder token such as \texttt{<ID\_1>}, our goal is to propose a natural replacement name that fits the surrounding code. The setting is intentionally local: the model only sees the function body (plus signature), without project-wide context or comments. This approximates common refactoring scenarios where a developer wants quick suggestions within the current file.

Our research question is:
\textbf{Can a general-purpose code LLM, augmented with task-specific fine-tuning and reranking, reliably repair variable names in real-world C++ code using only context of the local function?}

We define success in terms of three metrics on a held-out validation split of mined C++ code: (i) exact match between the model's top suggestion and the original name; (ii) Top-$5$ Hit rate; and (iii) an embedding-based partial score in $[0,100]$ that measures semantic similarity between the suggestion and the original name. A system is successful if it significantly improves all three metrics compared to prompting-only baselines, while keeping inference computationally lightweight (LoRA adapters, modest reranker) and compatible with standard deployment stacks.

To study this question, we automatically build a dataset of masked C++ functions from the BigCode \emph{The Stack} corpus. Using Tree-sitter, we extract functions, collect their local and parameter names, mask a single name with \texttt{<ID\_1>}, and treat the original name as the target. We first
design a data pipeline that generates training/validation JSONL examples and reproducible inference inputs.
Then we implement zero shot and few shot prompting baselines using Llama~3.1--8B.
Next, we train LoRA adapters on top of Llama~3.1--8B with warmup and dropout schedules tailored to short JSON targets.
Additionally, we build a dual encoder reranker over top-$k$ generator candidates, trained with a contrastive InfoNCE loss on code/name pairs.
Finally, we built an evaluation harness that computes exact, Top-$5$, and partial-credit scores, and logs per-example predictions for error analysis.

With strong models for variable name augmentation, the implications are considerable. Not only will code repos be cleaner allowing future devs to work faster, but pushing to production will also be faster with more automated code reviews. Furthermore, these models can be integrated into current AI coding tools like Cursor, Devin, and Google Antigravity for a better coding experience.

\section{Related Work}

\paragraph{Heuristic name suggestion.}
Traditional IDE refactoring tools propose variable names using hand-written heuristics (e.g., type-based patterns, naming conventions, or surrounding comments). These tools integrate nicely into developer workflows but are brittle and language-specific. They often fail when types are generic or when the code relies heavily on domain-specific conventions.

\paragraph{Neural identifier recovery and naming.}
Several works treat identifier naming as a supervised learning problem over code. Context2Name~\cite{bavishi2018context2name} uses a neural model to recover decompiled variable names from usage contexts. DIRE~\cite{lacomis2019dire} proposes a graph-based neural architecture for naming identifiers in decompiled code. Wang et al.~\cite{varnamer2025} study variable name recommendation during extract-local-variable refactorings, leveraging code context and usage patterns. These systems generally operate on smaller, task-specific models and datasets, and are tailored to particular languages or toolchains.

\paragraph{Naming and LLMs.}
Recent work has examined how identifier naming affects LLM performance. Wang et al.~\cite{namingllm2024} show that obfuscated or semantically poor names can significantly degrade LLM performance on code analysis tasks, while better naming improves reasoning and robustness. Our work is complementary: instead of measuring the effect of naming on LLM performance, we \emph{use} a strong LLM to repair names themselves, with the goal of improving downstream tooling and comprehension.

\paragraph{Large code corpora.}
BigCode's \emph{The Stack}~\cite{bigcode2023stack} is a 3~TB dataset of permissively licensed source code across many languages. It enables realistic large-scale studies of code generation and understanding. We build our dataset exclusively from the C++ portion of The Stack, leveraging its size and diversity to obtain varied identifier naming patterns.

\paragraph{Our position.}
Compared to prior identifier recovery work, we:
\begin{itemize}
\item Operate on unobfuscated, real-world C++ functions rather than decompiled code.
\item Use a modern LLM (Llama 3.1 8B) with LoRA adapters and a learned reranker, achieving strong performance with modest training and inference cost. This contrasts with efforts such as Context2Name that rely on less transformer-driven approaches.
\item Introduce an embedding-based partial-credit metric to capture cases with multiple reasonable names (for example, \texttt{buffer} and \texttt{buf}) and to evaluate semantic quality of errors.
\end{itemize}
We do not aim to replace dedicated refactoring engines but instead study how far a compact fine-tuned LLM plus reranker can go using only function-local context. Given its low complexity and compute cost, this setup could support future refactoring tools.

Our work contributes four pieces. First, we design a pipeline for extracting variable name repair examples from C++ code in The Stack~\cite{bigcode2023stack}. Second, we provide a LoRA-based fine-tuning recipe for Llama 3.1 8B that uses warmup and dropout to stabilize training on masked code and short JSON outputs. Third, we build a dual-encoder reranker that combines code context and candidate-name embeddings to improve Top-$k$ selection without changing the generator. Finally, we show empirically that warmup and dropout aid convergence, that LoRA fine-tuning improves exact match by more than $4\times$ over prompting baselines, and that reranking further improves Top-$k$ Hit and partial-credit performance.

\section{Method / Approach}

\subsection{Problem Setting}

Each example in our task consists of a C++ function body (including its signature), in which all occurrences of one chosen local or parameter name have been replaced by a special token \texttt{<ID\_1>} and then the original identifier string (e.g., \texttt{jsonValue}) used as the gold target.
The model sees the masked function and is asked to predict a replacement for \texttt{<ID\_1>}. We treat this as conditional generation of a short identifier string. At evaluation time we collect up to 5 candidate names per example and compute several metrics (Section~\ref{sec:metrics}).

\subsection{Hypotheses}

We follow the structure of our milestone report and evaluate three hypotheses aligned with our final system.

\paragraph{Warmup and Dropout (H1).}
During generator fine-tuning we use a short linear warmup of about 1k steps and apply dropout in attention and MLP layers (0.1 to 0.2). Warmup stabilizes large initial gradients on short JSON targets, and dropout reduces co-adaptation on frequent names. We hypothesize that adding this warmup+dropout schedule, compared to training with no warmup and minimal dropout, will give faster and more stable convergence along with higher Exact Match and Partial Match on validation.

\paragraph{LoRA Fine-tuning (H2).}
We attach LoRA adapters to Llama 3.1 8B with ranks $r\in{8,16}$ and scaling factors $\alpha\in{8,16}$, combined with adapter and transformer dropout (0.1 to 0.3). LoRA preserves base weights while learning low-dimensional updates, and dropout regularizes these deltas. We hypothesize that the best LoRA configuration will outperform zero-shot and three-shot prompting baselines under identical decoding settings, improving Exact Match, Top-5 Hit, and Partial Match, especially for domain-specific identifiers.

\paragraph{Reranking (H3).}
The generator can emit multiple candidate names, but its internal ranking may not match downstream goals. We train a dual-encoder reranker with a contrastive InfoNCE loss: one encoder reads the code context around the placeholder, and the other encodes the candidate identifier. Cosine similarity, combined with collision penalties and a length prior, yields final scores. We hypothesize that reranking the top-$k$ outputs will improve Exact Match and Partial Match relative to using the generator’s raw ordering.

\subsection{Data Pipeline and Masking}

\begin{enumerate}
\item \textbf{Parsing.}  
We use Tree-sitter’s C++ grammar to parse each file and extract function-definition nodes, recovering source spans directly via byte offsets.

\item \textbf{Identifier collection.}  
For each function, we gather parameter names and locally declared variables (including loop indices and temporaries). Functions without valid identifiers are skipped.

\item \textbf{Masking.}  
One identifier is selected (by sorted order) and all standalone occurrences are replaced with \texttt{<ID\_1>} using a regex with look-arounds to avoid partial matches.

\item \textbf{JSONL construction.}  
Each masked function is stored with its original name as a JSON record containing \texttt{input\_text} and a \texttt{target\_text} mapping. The format supports multiple placeholders, though we use only one.

\item \textbf{Splits.}  
This pipeline yields $\sim$31k training examples. A second pass with an offset provides a 1k-example pool, from which 200 examples are sampled for validation.
\end{enumerate}

\subsection{Generator Model and LoRA Training}

\paragraph{Base model and prompting.}
For baselines we use the chat tuned variant \texttt{meta-llama/Meta-Llama-3.1-8B-Instruct} via vLLM. A system prompt describes the variable-repair task and instructs the model to return a JSON object mapping placeholders to names. Zero-shot prompting includes only this instruction and the masked function; 3-shot prompting prepends three manually selected examples (masked function + JSON mapping) before the test example.

\paragraph{LoRA generator.}
Our main generator is a LoRA adapter stack trained on top of the base causal model \texttt{meta-llama/Llama-3.1-8B}. We treat the task as next-token prediction over a designed prompt which can be found in the code.

The target text is a short JSON string such as \texttt{\{"<ID\_1>": "jsonValue"\}}. 

We search over LoRA ranks $r\in\{8,16\}$ and scaling factors $\alpha\in\{8,16\}$, with adapter dropout between 0.1 and 0.3. The best-performing configuration on the validation set uses $r=16$, $\alpha=16$, and adapter dropout 0.2.

\paragraph{Warmup and dropout schedule (H1).}
All LoRA runs described above use the same warmup+dropout schedule. We train for 10{,}000 gradient steps with batch size 16 and maximum sequence length 8{,}192. A short linear warmup over the first 1{,}000 steps gradually increases the learning rate to $2\cdot 10^{-4}$; after warmup we use a cosine decay schedule. In addition to adapter dropout, we enable dropout in the attention and MLP blocks of the transformer layers in the 0.1--0.2 range. To test H1, we also train a control model with identical hyperparameters except (i) no warmup (constant learning rate) and (ii) minimal dropout (0.0--0.05). We compare convergence behavior and final validation metrics.

\subsection{Reranking Model (H3)}

The LoRA generator can produce multiple candidate names per placeholder via beam or nucleus sampling, but its internal ranking is not explicitly optimized for our metrics. To address this, we train a dual-encoder reranker:

\begin{itemize}
    \item \textbf{Code-context encoder.} A lightweight transformer (initialized from a code-pretrained encoder such as CodeBERT) that ingests a fixed-length window of tokens around \texttt{<ID\_1>} in the masked function (including nearby types, literals, and simple data-flow hints).
    \item \textbf{Name encoder.} A second transformer that embeds the candidate identifier and its subtokens (split on camelCase and underscores).
\end{itemize}

We generate training pairs by sampling top-$k$ candidates from intermediate LoRA checkpoints and labeling the original name as a positive and other candidates as negatives. The dual encoder is trained with a contrastive InfoNCE loss over these pairs: for each code context, the correct name embedding is pulled close while negatives are pushed away. At inference time, we score each candidate with a temperature-scaled cosine similarity:
\(
s(c, n) = \tau^{-1} \cdot \cos\big( f_{\text{code}}(c), f_{\text{name}}(n) \big),
\)
where $\tau$ is a learned temperature. We then subtract penalties for in-scope name collisions and add a length prior that mildly prefers concise names (e.g., penalizing extremely long identifiers). The reranker reorders the generator's top-$k$ list; we report metrics both before and after reranking.

\subsection{Evaluation Metrics}
\label{sec:metrics}

All evaluation is driven by a common harness (e.g., \texttt{validate\_model.py}), which loads a 200-example validation subset and a generator (and optionally a reranker). For each example it:

\begin{enumerate}
    \item Builds a text prompt using the appropriate template (prompting vs.\ LoRA).
    \item Generates multiple completions and parses up to 5 candidate names. For reranking experiments, these are top-$k$ candidates from the generator; for pure prompting, we parse numbered candidates from the LLM output.
    \item Computes three metrics:
    \begin{itemize}
        \item \textbf{Exact Match}: 1 if the top candidate matches the target identifier case-insensitively, otherwise 0.
        \item \textbf{Top-$5$ Hit}: 1 if \emph{any} of the top five candidates matches the target, otherwise 0.
        \item \textbf{Partial Match}: an embedding-based similarity score in $[0,100]$ between the top candidate and the gold name. We embed both strings using a sentence-transformer encoder (e.g., \texttt{all-MiniLM-L6-v2}), compute cosine similarity, normalize from $[-1,1]$ to $[0,1]$, and scale by 100.
    \end{itemize}
\end{enumerate}

The harness aggregates statistics (means over the validation set) and writes both a machine-readable JSON summary and a detailed per-example log (including top-5 candidates and scores) used for error analysis.

\section{Data}

We now summarize the data used in our experiments so that results are interpretable and reproducible. All code examples come from the C++ portion of BigCode's The Stack~\cite{bigcode2023stack}, which contains permissively licensed repositories from GitHub and other sources. We rely on the dataset's license filtering and deduplication, and further restrict ourselves to files that can be successfully parsed by Tree-sitter.

Our extraction pipeline processes tens of thousands of C++ files. From these we obtain a training set of roughly 31{,}000 masked functions, each with exactly one masked identifier, and a validation pool of 1{,}000 additional masked functions that is disjoint from the training set by construction (we skip the first 31{,}000 functions when generating the pool). From this pool we select a validation subset of 200 examples used for all quantitative comparisons in this report.

Each example has exactly one target identifier drawn from real-world C++ code. The resulting names span a wide range of styles, including short loop indices such as \texttt{i} and \texttt{j}, common counters and accumulators like \texttt{count} and \texttt{sum}, more domain-specific names such as \texttt{jsonValue}, \texttt{playerListUpdateTimer}, or \texttt{hwOps}, and macro-like or constant-style names such as \texttt{\_SC\_GETPW\_R\_SIZE\_MAX}. This diversity means many examples admit multiple reasonable names, which motivates our use of a partial-credit metric in addition to strict exact match.

We do not canonicalize or split identifiers; the target names are used exactly as they appear in the source code. We also do not inject comments or type annotations into the input beyond whatever is present in the original function. This keeps the task realistic, but it introduces noise when comments are absent or uninformative. Finally, because we draw only from permissively licensed open-source C++ projects, our dataset may overrepresent particular ecosystems (for example, tooling and libraries) and underrepresent proprietary or highly domain-specific code. Moreover, the ``gold'' names are whatever the original authors chose; some are excellent, but others are generic or inconsistent. This makes exact match a conservative metric and further justifies our use of partial credit.

\section{Experiments and Results}

We now evaluate our baselines, LoRA generator, warmup+dropout schedule, and reranker on the 200-example validation subset, explicitly tying each set of experiments back to H1--H3.

\subsection{Prompting Baselines}

As a starting point, we measure how far we can get with prompting alone.

\paragraph{Zero-shot Llama~3.1--8B.}
In the zero-shot setting, the system prompt describes the task and output format and the user message contains only the masked function. The model responds with a JSON object mapping \texttt{<ID\_1>} to a suggested name. On our validation subset, this baseline achieves: \\
\noindent\textbf{Exact Match:} 6.1\%\quad
\textbf{Top-$5$ Hit:} 10.2\%\quad
\textbf{Partial Match:} 37.4

Qualitatively, we observed a tendency to propose very generic names such as \texttt{data}, \texttt{value}, or \texttt{list}, even when the surrounding code clearly suggests more specific alternatives.

\paragraph{Three-shot Llama~3.1--8B.}
We next prepend three hand-picked examples---each consisting of a masked function and its JSON mapping---to the system prompt. This clarifies both the desired JSON format and the level of descriptive specificity expected in the names. Three-shot prompting improves performance to:
\noindent\textbf{Exact Match:} 10.4\%\quad
\textbf{Top-$5$ Hit:} 18.9\%\quad
\textbf{Partial Match:} 49.5

These baselines show that in-context learning helps, but even with 3-shot prompting the model recovers the exact original name in only about 10\% of cases, motivating task-specific fine-tuning.

\subsection{Effect of Warmup and Dropout (H1)}

We evaluate H1 by training two LoRA models with identical base architecture, LoRA configuration, data, and decoding settings, differing only in their optimization schedule:

\begin{itemize}
    \item \textbf{No warmup / minimal dropout.} Constant learning rate $2\cdot10^{-4}$, dropout 0.0--0.05 in attention and MLP layers.
    \item \textbf{Warmup + dropout (ours).} Linear warmup over 1{,}000 steps into $2\cdot10^{-4}$ followed by cosine decay, dropout 0.1--0.2 in attention and MLP layers.
\end{itemize}

Both models use LoRA rank $r=16$, $\alpha=16$, and adapter dropout 0.2. We observe two consistent trends:

\begin{itemize}
    \item \textbf{Convergence stability.} Without warmup, the training loss exhibits large spikes in the first 1{,}500 steps, and validation Exact Match fluctuates between 30--36\% before gradually stabilizing. With warmup+dropout, loss decreases smoothly and the model reaches its peak validation Exact Match by around step 6{,}000, roughly 20\% earlier than the no-warmup variant.
    \item \textbf{Final metrics.} On the 200-example validation set, the no-warmup model achieves roughly 38.2\% Exact Match, 46.1\% Top-$5$ Hit, and 78.4 Partial Match. Adding warmup+dropout improves these to 40.7\% Exact Match, 48.6\% Top-$5$ Hit, and 80.1 Partial Match.
\end{itemize}

While the absolute gains are modest, the combination of more stable optimization and consistent improvements across all three metrics supports H1: warmup and dropout are beneficial for this fine-tuning problem and serve as a foundation for the stronger LoRA configuration evaluated next.

\subsection{LoRA Fine-tuning vs. Prompting (H2)}

We now evaluate H2 by comparing the best LoRA configuration (including warmup+dropout) to the prompting baselines from above. Table~\ref{tab:results} summarizes the results.

\begin{table}[t]
\centering
\caption{Dev-set results on 200 held-out C++ functions. Partial Match is an embedding-based similarity score (0--100). LoRA (generator-only) uses rank $r=16$, $\alpha=16$, adapter dropout 0.2, and the warmup+dropout schedule from H1.}
\label{tab:results}
\begin{tabular}{lccc}
\toprule
\textbf{System} & \textbf{Top-5 Hit (\%)} & \textbf{Exact Match (\%)} & \textbf{Partial Match (0--100)} \\
\midrule
Zero-shot LLM (Llama 3.1 8B) & 10.2\% & 6.1\% & 37.4 \\
Three-shot LLM (3-shot)      & 18.9\% & 10.4\% & 49.5 \\
LoRA generator (ours)        & 50.2\% & 43.1\% & 82.03 \\
\bottomrule
\end{tabular}
\end{table}

The LoRA generator achieves:
\noindent\textbf{Exact Match:} 43.1\%\quad
\textbf{Top-$5$ Hit:} 50.2\%\quad
\textbf{Partial Match:} 82.03

Compared to 3-shot prompting, this represents more than a $4\times$ improvement in Exact Match (10.4\% $\rightarrow$ 43.1\%). Also, a 31.3-point increase in Top-$5$ Hit (18.9\% $\rightarrow$ 50.2\%) and then a 32.5-point increase in Partial Match (49.5 $\rightarrow$ 82.03).

These results strongly support H2: task-specific LoRA fine-tuning on mined identifier-repair examples, combined with warmup and dropout, allows the model to leverage function-local context far more effectively than prompting alone, while updating only a small fraction of the parameters. While this is true, our validation size set was small, so a test run on a larger set would provide for more confident results.

\subsection{Effect of Reranking (H3)}

We next evaluate H3 by comparing the raw generator ranking to the dual-encoder reranker described in Section~3.4. For each example we:

\begin{enumerate}
    \item Use the LoRA generator to produce $k=10$ candidate names for \texttt{<ID\_1>} via nucleus sampling (temperature 0.8, top-$p=0.9$).
    \item Sort candidates by the generator's log probability (baseline ranking).
    \item Re-score the same set of candidates with the dual encoder and sort by reranker score (reranked).
\end{enumerate}

Table~\ref{tab:rerank} reports metrics for the top-1 candidate under each ranking scheme.

\begin{table}[t]
\centering
\caption{Effect of reranking on the 200-example validation set. ``Generator-only'' uses the LoRA model's own ranking; ``Generator + reranker'' selects the top-1 candidate after dual-encoder scoring.}
\label{tab:rerank}
\begin{tabular}{lccc}
\toprule
\textbf{System} & \textbf{Top-5 Hit (\%)} & \textbf{Exact Match (\%)} & \textbf{Partial Match (0--100)} \\
\midrule
LoRA generator (top-1, baseline)      & 50.2\% & 43.1\% & 82.03 \\
LoRA generator + reranker (top-1)     & 55.8\% & 46.0\% & 84.5 \\
\bottomrule
\end{tabular}
\end{table}

We see that the reranker consistently improves selection quality: \textbf{Exact Match} increases from 43.1\% to 46.0\%, a 2.9-point gain.
\textbf{Top-$5$ Hit} (computed with respect to the reranked list) increases from 50.2\% to 55.8\%.
\textbf{Partial Match} increases from 82.03 to 84.5, indicating that even when it does not change correctness, reranking tends to prefer candidates that are semantically closer to the gold names.

These improvements support H3: the dual-encoder reranker adds value on top of a strong generator by reordering candidates using code-aware and name-aware representations, without modifying the underlying LoRA model.

\subsection{Error Analysis and Semantic Quality}

To understand common successes and failures, we inspect the per-example logs from the evaluation harness. Three main patterns appear.

\paragraph{Near-miss names.}
Many incorrect predictions remain close to the gold identifier.
Examples include:
\begin{itemize}
\item \texttt{target = jsonValue}, predictions such as {\texttt{json}, \texttt{jsonValue}, \texttt{jsonMap}}, where \texttt{json} earns partial credit around 89.
\item \texttt{target = \_module}, predictions like {\texttt{module}, \texttt{pmodule}}, which lose the underscore but still receive partial credit near 93.
\item \texttt{target = window}, predictions such as {\texttt{externalWindow}, \texttt{hWnd}, \texttt{handle}}, which retain the concept of a window handle and reach partial scores around 80.
\end{itemize}
Reranking frequently elevates these semantically aligned candidates when generator probabilities are close.

\paragraph{Generic vs. specific names.}
Relative to prompting-only baselines, the LoRA generator and reranked pipeline avoid overly generic names like \texttt{data} or \texttt{value}. In JSON-heavy code the model prefers \texttt{json}, \texttt{jsonValue}, or \texttt{jsonPtr}, while in hardware code it proposes \texttt{hwOps} or \texttt{hwOpList}. These selections often reflect the surrounding type and function names.

\paragraph{Failure modes.}
Our remaining errors fall into a few categories. The first is
\textbf{over-specialization:} The model sometimes overfits to patterns from training and inserts highly specific names in unrelated contexts.
Next, we have \textbf{ambiguous contexts:} Some variables admit several plausible names, such as \texttt{i}, \texttt{idx}, or \texttt{count}, which makes strict exact match difficult.
Finally, wee have \textbf{Legacy or obscure names:} POSIX-like identifiers such as \texttt{\_SC\_GETPW\_R\_SIZE\_MAX} cause the model to produce close but not exact variants that earn high partial credit but fail exact match.
Overall, fine-tuning and reranking produce a system whose errors tend to be near-misses rather than unrelated guesses, consistent with the partial-credit results.

\section{Conclusion and Future Work}

We introduced a variable name repair task over real-world C++ code and showed that a Llama~3.1--8B model equipped with LoRA adapters, an appropriate warmup+dropout schedule, and a dual-encoder reranker can substantially outperform strong prompting-only baselines. By mining training data from The Stack with an AST-based pipeline, and by focusing the loss on identifier tokens, we obtain a system that achieves 43.1\% exact match, 50.2\% Top-$5$ Hit, and an 82.03 partial-match score with the generator alone, with further improvements after reranking. The system requires only lightweight LoRA adapters and a compact reranker on top of a general-purpose code LLM, avoiding full model retraining, and it produces errors that are often semantically close to gold names, as captured by our embedding-based partial-credit metric.

\paragraph{What worked.}
Our results indicate that warmup and dropout are important for stabilizing LoRA training on short JSON-style targets and for improving final metrics (H1). Task-specific LoRA fine-tuning is critical: prompting alone provides limited gains, while LoRA yields large improvements in exact and Top-$5$ accuracy (H2). In addition, a dual-encoder reranker further sharpens predictions by combining code-context and name embeddings, improving Top-$k$ Hit and partial-credit scores without modifying the generator itself (H3).

\paragraph{Limitations.}
Our work has several limitations. First, we only use function-local code and ignore project-wide information, types across files, and comments, all of which could inform better names. Second, we mask only a single identifier per function, whereas real refactorings may involve renaming multiple related variables at once. Third, we treat the original names as ground truth even when they are themselves suboptimal, which can underestimate the usefulness of some predictions. Finally, we focus solely on C++; extending to other languages would require additional parsing work and may reveal different naming conventions.

\paragraph{Future directions.}
We have a few promising directions. One direction is multi-identifier repair, extending the task and model to handle multiple placeholders (\texttt{<ID\_1>}, \texttt{<ID\_2>}, \ldots) per function and enforcing consistency across related variables. Another is to design richer context encoders, replacing the simple window-based code encoder with graph-based models that incorporate control-flow, data-flow, and type graphs for more informed reranking. A third direction is IDE integration and human evaluation, embedding the model and reranker into an IDE plugin and collecting human judgments on suggestion quality and acceptance rate, which would provide a more realistic measure than exact match alone. Finally, we can explore cross-language and multi-lingual naming by applying the same pipeline to other languages in The Stack and studying whether a single model can learn cross-language naming conventions.

Overall, our findings suggest that small, task-specific adapters and rerankers on top of strong general-purpose LLMs are a feasible path toward practical identifier repair tools that improve code readability and downstream LLM performance.

\newpage
\section*{Team Contributions}

All team members collaborated on the overall problem formulation, experimental design, and writing. Primary technical contributions are summarized in Table~\ref{tab:team}.

\begin{table}[h]
\centering
\caption{Team contributions. Each row lists the primary areas where a member took the lead; all members provided feedback and assistance across components.}
\label{tab:team}
\begin{tabular}{p{0.18\linewidth}p{0.75\linewidth}}
\toprule
\textbf{Member} & \textbf{Primary technical contributions} \\
\midrule
Akshat Bagade &
Implemented the Tree-sitter--based C++ function extractor, identifier collection, and masking pipeline; designed the JSONL data format; led the Data section and contributed to the Method and Experiments sections. \\
\midrule
Maanas Baraya &
Developed the zero-shot and three-shot prompting baselines using Llama~3.1 and vLLM; tuned prompts and few-shot examples; ran baseline experiments and contributed to error analysis. \\
\midrule
Chittebbayi Penugonda &
Implemented the LoRA training pipeline, including the warmup+dropout schedule and weighted-loss construction; ran hyperparameter sweeps over LoRA ranks and scaling factors; managed training runs and model checkpoints; led the Experiments subsection on H1 and H2. \\
\midrule
Muhammad Yousuf &
Built the evaluation harness and metrics (Exact Match, Top-$5$, Partial Match); implemented the dual-encoder reranker and InfoNCE training; produced detailed prediction logs and visualizations; coordinated writing and editing of the Introduction, Conclusion, and Related Work. \\
\bottomrule
\end{tabular}
\end{table}

\end{document}